\documentclass{svjour3}                    % onecolumn
\smartqed  % flush right qed marks, e.g. at end of proof
\usepackage{graphicx}
\usepackage{mathptmx}      % use Times fonts if available on your TeX
\usepackage{amsmath,verbatim}
\usepackage[pdftex,usenames]{color}
\usepackage{hyperref}
\usepackage{rotating}
\usepackage{multirow}
\usepackage[english]{babel}
\usepackage{microtype}
\usepackage{comment}
\usepackage[normalem]{ulem}
\usepackage{rotating}
\usepackage{lipsum}
\usepackage{latexsym}
\newcommand{\av}[1]{\langle {#1} \rangle}

\newcommand{\km}{q_\mathrm{max}}

\journalname{J Stat Phys}
\begin{document}

\title{Eigenvector localization in real networks and its implications for
  epidemic spreading
  \thanks{We acknowledge financial support from the Spanish MINECO,
    under Projects No. FIS2013-47282-C2-2 and
    No. FIS2016-76830-C2-1-P. R. P.-S. acknowledges additional financial
    support from ICREA Academia, funded by the Generalitat de
    Catalunya.}  }

\subtitle{}

\titlerunning{Eigenvector localization in
  real-networks} % if too long for running head

\author{Romualdo Pastor-Satorras \and Claudio Castellano}

\authorrunning{R. Pastor-Satorras and C. Castellano} % if too long for running head

\institute{Romualdo Pastor-Satorras \at 
           Departament de F\'\i sica i Enginyeria Nuclear,
           Universitat Polit\`ecnica
           de Catalunya, Campus Nord B4, 08034
           Barcelona, Spain \and
           Claudio Castellano   \at
           Istituto dei Sistemi Complessi (ISC-CNR), via dei Taurini
           19, I-00185 Roma, Italy \at
           Dipartimento di Fisica, ``Sapienza'' Universit\`a di
           Roma, P.le A. Moro 2, I-00185 Roma, Italy}

\date{Received: date / Accepted: date}

\maketitle

\begin{abstract}
  
  The spectral properties of the adjacency matrix, in particular its
  largest eigenvalue and the associated principal eigenvector, dominate many
  structural and dynamical properties of complex networks. Here we focus
  on the localization properties of the principal eigenvector in real
  networks. We show that in
  most cases it is either localized on the star defined by the node with
  largest degree (hub) and its nearest neighbors, or on the densely
  connected subgraph defined by the maximum $K$-core in a $K$-core
  decomposition. The localization of the principal eigenvector is often
  strongly correlated with the value of the largest eigenvalue, which is
  given by the local eigenvalue of the corresponding localization
  subgraph, but different scenarios sometimes occur.
  We additionally show that simple targeted
  immunization strategies for epidemic spreading are extremely sensitive
  to the actual localization set.
    

\end{abstract}

\section{Introduction}
\label{intro}

The adjacency matrix encodes all the information about a complex
network~\cite{Newman10} and, as a consequence, about its structure and
function.  The knowledge of its spectral properties provides a
fundamental tool to understand the network topology and the behavior of
dynamical processes mediated by it~\cite{barratbook}.  In particular the
largest eigenvalue (LEV) $\Lambda_M$ of the adjacency matrix determines
the position of critical transitions for processes as diverse as
epidemics~\cite{Chakrabarti_2008,Castellano2010,Goltsev2012},
synchronization~\cite{Restrepo2005}, weighted percolation on directed
networks~\cite{Restrepo2008}, models of genetic
control~\cite{Pomerance2009}, or the dynamics of excitable
elements~\cite{Kinouchi:2006aa}. On the other hand, the principal
eigenvector (PEV), $\{ f_i\}$, corresponding to the LEV, is at the core
of a set of centrality measures, such as Katz's
centrality~\cite{Katz1953} or PageRank~\cite{Brin1998}, and it has been
associated with the validity of mean-field approaches for the study of
epidemic spreading in networks~\cite{Goltsev2012}.

A fundamental problem in network science is hence the understanding of
how $\Lambda_M$ and the structure of the PEV depend on network
features. Given the adjacency matrix, the determination of the largest
eigenpair is numerically not difficult, by means of the straightforward
power-iteration method~\cite{golub2012matrix}, even for the largest
structures. However, understanding what determines the largest eigenpair
is a very important task, since it allows us to efficiently
control and manipulate spectral properties and the associated dynamical
effects~\cite{Restrepo2006,Milanese2010,VanMieghem2011}.

A first, seminal result in this endeavor was provided by Chung, Lu and
Vu~\cite{Chung03} (CLV), who derived, for a class of maximally
random networks with given expected degree distribution, a formula
relating the LEV with simple topological properties of the network,
namely the first moments of the degree distribution and the overall
largest degree, $\km$.  In the limit of networks of infinite size the
CLV expression can be recast as~\cite{Castellano2010}
\begin{equation}
  \label{eq:1}
  \Lambda_M \approx \max \{ \sqrt{\km}, \av{q^2}/\av{q}\},
\end{equation}
which turns out to be remarkably accurate for random uncorrelated
networks of any size~\cite{Castellano2017}, as generated with the
uncorrelated configuration model (UCM) \cite{Catanzaro05}.
Eq.~(\ref{eq:1}) can be physically interpreted as
follows~\cite{Castellano2012}: the largest eigenvalue of the whole
structure is given by the largest among the eigenvalues of two network
subgraphs, considered as isolated. The first of these subgraphs is the
star graph formed by the node with largest degree (the hub) and its
nearest neighbors, with a LEV
$\Lambda_M^{(star)} = \sqrt{\km}$~\cite{PVM_graphspectra}.  The second
is the densely interconnected subgraph identified, in the $K$-core
decomposition procedure~\cite{Seidman1983269}, as the $K$-core with
maximum index $K_M$.  This subgraph, being homogeneous, has a LEV well
approximated by its internal average degree,
$\Lambda_M^{(K_M)} \approx \av{q}_{K_M}$.  In
Ref.~\cite{Dorogovtsev2006} it is shown that the internal average degree
of the maximum $K$-core can be approximated by the excess average degree
of the whole network, $\av{q^2}/\av{q}$. Therefore, we have
$\Lambda_M^{(K_M)} \approx \av{q^2}/\av{q} $.  This interpretation is
validated by the observation that, in random uncorrelated networks, when
the global LEV is given by $\sqrt{\km}$, the PEV is localized on the
star graph composed by the hub and its neighbors; in the other case the
PEV is instead localized on the max $K$-core~\cite{Pastor-Satorras2016}
.

While Eq.~(\ref{eq:1}) predicts very well the LEV for random
uncorrelated networks, it often fails when applied to real-world
networks~\cite{Castellano2017}.  
Recently, the original CLV formula has been modified to
extend its validity beyond the uncorrelated case~\cite{Castellano2017}:
\begin{equation}
  \Lambda_M \approx \max \{\sqrt{\km},\av{q}_{K_M}\},
  \label{conjecture}
\end{equation}
where $\av{q}_{K_M}$ is the average degree of the max $K$-core subgraph.
The physics behind Eq.~(\ref{conjecture}) is the same as in the
uncorrelated case: the value of the LEV is determined by the competition
among two subgraphs, the star centered around the hub and the max
$K$-core.  What changes with respect to the uncorrelated case is the
expression of $\Lambda_M^{(K_M)}$, which may largely differ from the
ratio between the second and the first moment of the degree
distribution.  Eq.~(\ref{conjecture}) has been empirically tested on a
large set of real-world networks, for which a good accuracy in the
prediction of the actual value of $\Lambda_M$ was
observed~\cite{Castellano2017}.

In this paper we push further the analysis of
Ref.~\cite{Castellano2017}, by studying in detail the localization
properties of the PEV in real networks, spelling out some of its
implications and checking their validity in real-world networks.  After
reviewing established results for synthetic networks, we analyze a set
of 38 real-world topologies of diverse origin, with the goal of
determining whether the dominating term in Eq.~(\ref{conjecture}) allows
us to infer where the PEV is localized. While this picture is most
of the time correct, we uncover the existence of some
exceptions: networks with the PEV localized on both subgraphs and (more
surprisingly) networks with the PEV localized on a set of nodes with no
overlap with the star surrounding the hub or the max $K$-core.  In
addition, we investigate how the PEV localization nontrivially affects a
dynamical (epidemic) process mediated by the network topology.  The
removal of the hub or of the max $K$-core can lead to negligible effects
or to a complete disruption of the epidemic dynamics, depending on the
PEV localization properties.

\section{Localization on subgraphs in synthetic networks}
\label{sec:local-synth-netw}

We first consider the localization of the PEV on synthetic uncorrelated
networks with a power-law degree distribution of the form
$P(q) \sim q^{-\gamma}$, generated using the uncorrelated configuration
model (UCM)~\cite{Catanzaro05}. For the PEV $\{ f_i\}$, normalized as
$\sum_i f_i^2 = 1$, the localization in ensembles of synthetic networks
of varying size can be assessed by studying the inverse participation ratio
(IPR) $Y_4$, defined as
\begin{equation}
  \label{eq:3}
  Y_4(N) = \sum_i f_i^4,
\end{equation}
as a function of the network size $N$~\cite{Goltsev2012,2014arXiv1401.5093M}.
Different types of localization reflect in the functional form of the
IPR as a function of $N$, expressed as a
power-law~\cite{Pastor-Satorras2016}:
\begin{equation}
  \label{eq:4}
  Y_4(N) \sim N^{-\alpha}.
\end{equation}
If the PEV is delocalized, i.e., homogeneously distributed over all nodes
in the network, then $f_i \simeq N^{-1/2}$, $\forall i$. In this case
$Y_4(N) \sim N^{-1}$ and $\alpha = 1$. If the PEV is instead
homogeneously localized on a set of nodes $V$ of size $N_v$, then
$f_i \simeq N_V^{-1/2}$ for $i \in V$, and $f_i \simeq 0$ for
$i \not\in V$. In this case, we have $Y_4(N) \sim N_V^{-1}$. If $V$ is a
subextensive set of nodes, with size $N_V$ growing with the size $N$ of
the whole network as $N_V \sim N^\delta$, then $\alpha = \delta$. In the
extreme case of localization of a set of nodes with fixed size,
$N_V = \mathrm{const.}$ (including the case of a single node), then
$\alpha = 0$.  Other more complex forms of localization are possible as
well.  One that plays a relevant role in this work is the one occurring
on a star graph formed by a hub connected to $\km$ nodes (leaves). In
such a case, it is easy to see, applying Perron-Frobenius
theorem~\cite{Gantmacher}, that the LEV is $\Lambda_M = \sqrt{\km}$, and
that the hub has weight $f_{hub} = 1/\sqrt{2}$, while each of the leaves
carries a weight $f_i=1/\sqrt{2\km}$. As a consequence the total weight
is equally split among the hub and the set of all leaves.  In this case,
$Y_4(\km) = \frac{1}{4} (1 - \km^{-1})$, and the IPR goes asymptotically
to $1/4$ as $\km$ diverges.

\begin{figure}[t]
  \begin{center}
    \includegraphics [width=\textwidth]{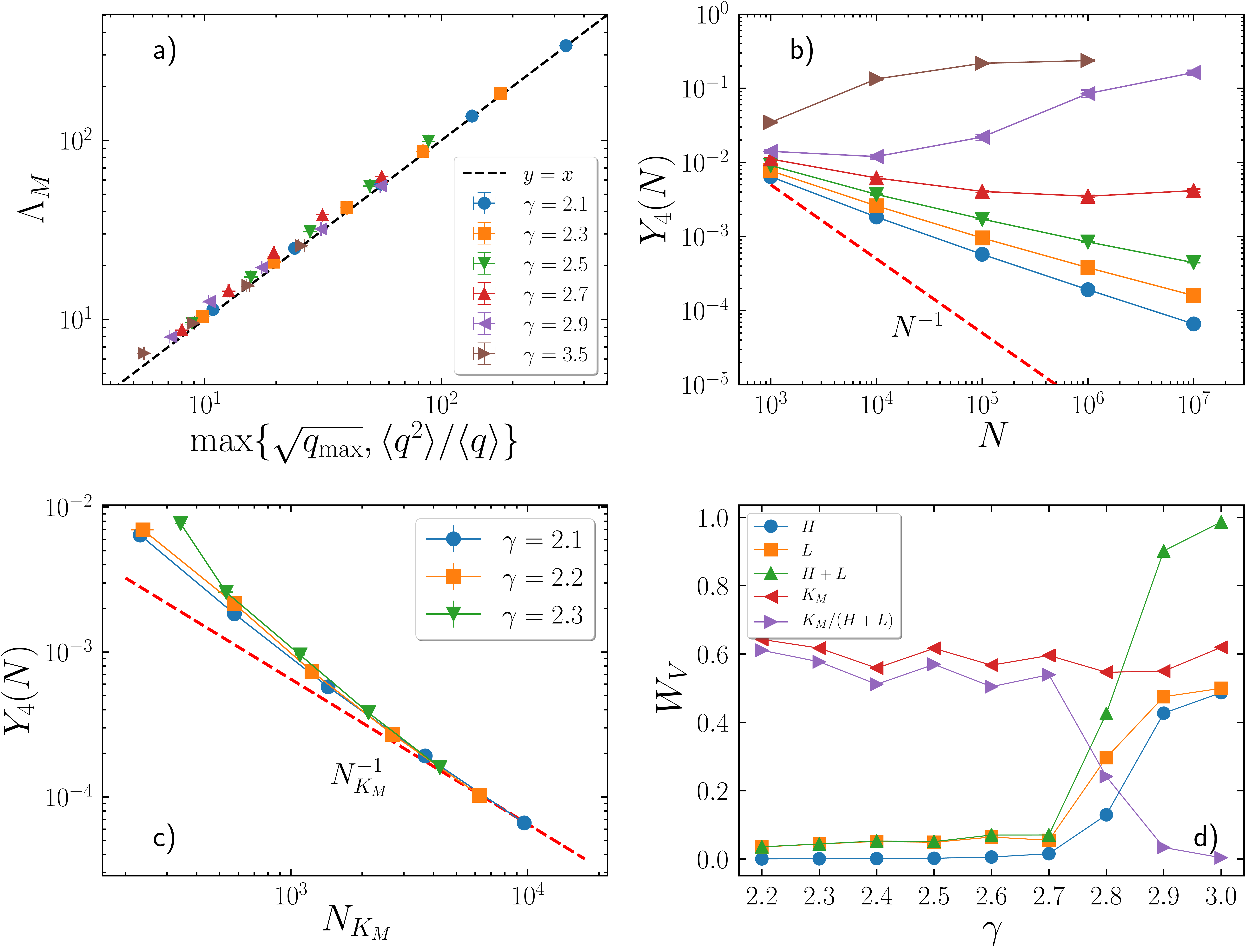}
    \caption{Spectral properties of uncorrelated UCM networks. (a)
      Largest eigenvalue $\Lambda_M$ as a function of the theoretical
      prediction Eq.~(\ref{eq:1}) for networks sizes from $N=10^3$ to
      $10^7$ (b) Inverse participation ratio $Y_4$ as a function of
      network size. (c) Inverse participation ratio as a function of the
      size of the max $K$-core $N_{K_M}$. (d) Relative weight $W_V$ of
      different network subgraphs $V$ (see main text). Network size is
      in this case $N=10^7$.}
      \label{UCM}
  \end{center}
\end{figure}

In Fig.~\ref{UCM}a) we first check the validity of the original CLV
result, Eq.~(\ref{eq:1}), by plotting the largest eigenvalue $\Lambda_M$
as a function of $\max \{ \sqrt{\km}, \av{q^2}/\av{q}\}$. In order to
avoid complications, we consider only networks with a single node with
maximum degree $\km$.  As we can observe, Eq.~(\ref{eq:1}) fits the
empirical data with an excellent precision. Given the form of
Eq.~(\ref{eq:1}), for a power-law degree distribution
$P(q) \sim q^{-\gamma}$ with $\gamma \leq 5/2$
the LEV is given by $\av{q^2}/\av{q}$, and the PEV is expected to be
localized on the max $K$-core. For $\gamma > 5/2$, on the other hand,
the LEV takes the form $\sqrt{\km}$, and the PEV is localized on the
hub. In Fig.~\ref{UCM}b) we check this conjecture by plotting the IPR as
a function of $N$. In accordance with our
expectation, for large $\gamma$ the value of $Y_4(N)$ tends to a
constant for large $N$, which is compatible with localization on the
hub and its neighbors, while for small $\gamma$, it decreases with $N$. 
The decrease for $\gamma \leq 5/2$ is however slower than $N^{-1}$, 
indicating the absence of complete delocalization~\cite{Pastor-Satorras2016}. 
The PEV is in this case localized on
a subextensive set, which is identified as the max $K$-core in
Fig.~\ref{UCM}c), where we plot $Y_4(N)$ as a function of the max
$K$-core size, and observe a trend $Y_4 \sim N_{K_M}^{-1}$ for large
$N$.

A correct characterization of the PEV localization in UCM networks is
somewhat impeded by the fact that the hub and some of its neighbors
usually belong to the max $K$-core. In order to properly discriminate
the weight contribution of each one of the relevant subgraphs, in
Fig.~\ref{UCM}d) we plot the relative weight
\begin{equation}
  \label{eq:5}
  W_V = \sum_{i \in V} f_i^2
\end{equation}
for different sets of nodes: the hub alone, $V=H$; the set of hub neighbors
(leaves), $V = L$; the star defined by the hub and the
leaves, $V = H + L$; the complete max $K$-core, $V = K_M$; and the max
$K$-core, excluding hubs and leaves, $V = K_M \setminus (H + L)$. 
As we can observe, for small $\gamma$ the weight is dominated by the max
$K$-core, with a very small contribution from the leaves and the hub
inside it.  For large $\gamma$, on the other hand, the contribution of
the hub and the leaves increases, in such a way as to almost dominate
the whole PEV weight for the largest $\gamma$ considered.  In this
regime, however, the max $K$-core still represents a sizeable fraction
of the total weight.  This is due to the fact that, even though in the
limit $\gamma \to 3$ the max $K$-core diminishes in
size~\cite{Dorogovtsev2006}, it always contains the hub and a fraction
of the leaves, which are the responsible for its large weight.  When
subtracting the weight of the hub and of the leaves it contains, the remaining
max $K$-core shows a vanishing weight. These observations indicate that,
for large $\gamma$, we are in the presence of a localization on the star
surrounding the hub.  The weight is equally divided between the hub and
the $\km$ leaves, i.e., $f_i^2 \sim 1/2$, $i \in H$, and
$f_i^2 \sim 1/(2\km)$, $i \in L$. The rest of the network has a
vanishing weight.  In this case, $Y_4(N) \sim (1/2)^2 + (1/2)^2/\km$,
which still goes to a constant for sufficiently large $N$, in agreement
with the result in Fig.~\ref{UCM}b).

\section{Localization on subgraphs in real networks}
\label{Localization}

\begin{figure}[t]
  \begin{center}
    \includegraphics [width=\textwidth]{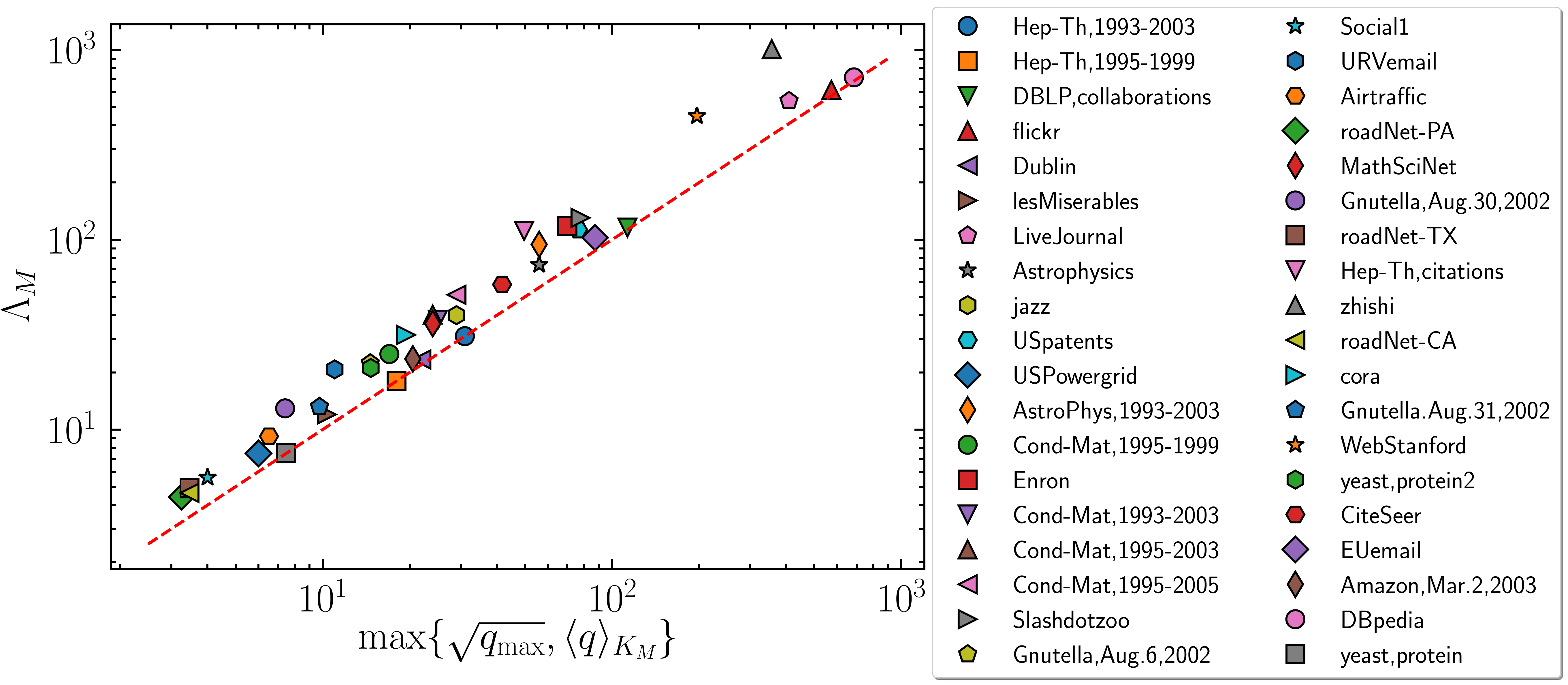}
    \caption{Largest eigenvalue $\Lambda_M$ as a function of the
      theoretical prediction Eq.~(\ref{conjecture}) for the 38 networks
      real networks listed in Table~\ref{tab:realnetworks}.}
      \label{LambdaM}
  \end{center}
\end{figure}

In analogy with the case of uncorrelated networks, it is very natural to
conjecture, from Eq.~(\ref{conjecture}), that when $\Lambda_M$
practically coincides with $\Lambda_M^{(star)}$ then the PEV is
localized on the hub and its immediate neighbors; vice versa, when
$\Lambda_M \approx \Lambda_M^{(K_M)}$ the PEV is localized on the max
$K$-core.  To test this hypothesis we consider a subset of the real
networks analyzed in Ref.~\cite{Castellano2017} given by those networks
in which the hub does not belong to the max $K$-core. We do not consider
networks in which the hub belongs to the max $K$-core since in that case
it is difficult to discriminate between a PEV localized only on the hub
and its neighbors or on all the max $K$-core.  This criterion leaves us
with 38 topologies, listed in Table~\ref{tab:realnetworks} (see
Appendix).  The validity of Eq.~(\ref{conjecture}) for those networks is
illustrated in Fig.~\ref{LambdaM}.

\begin{figure}[t]
  \begin{center}
    \includegraphics [width=0.9\textwidth]{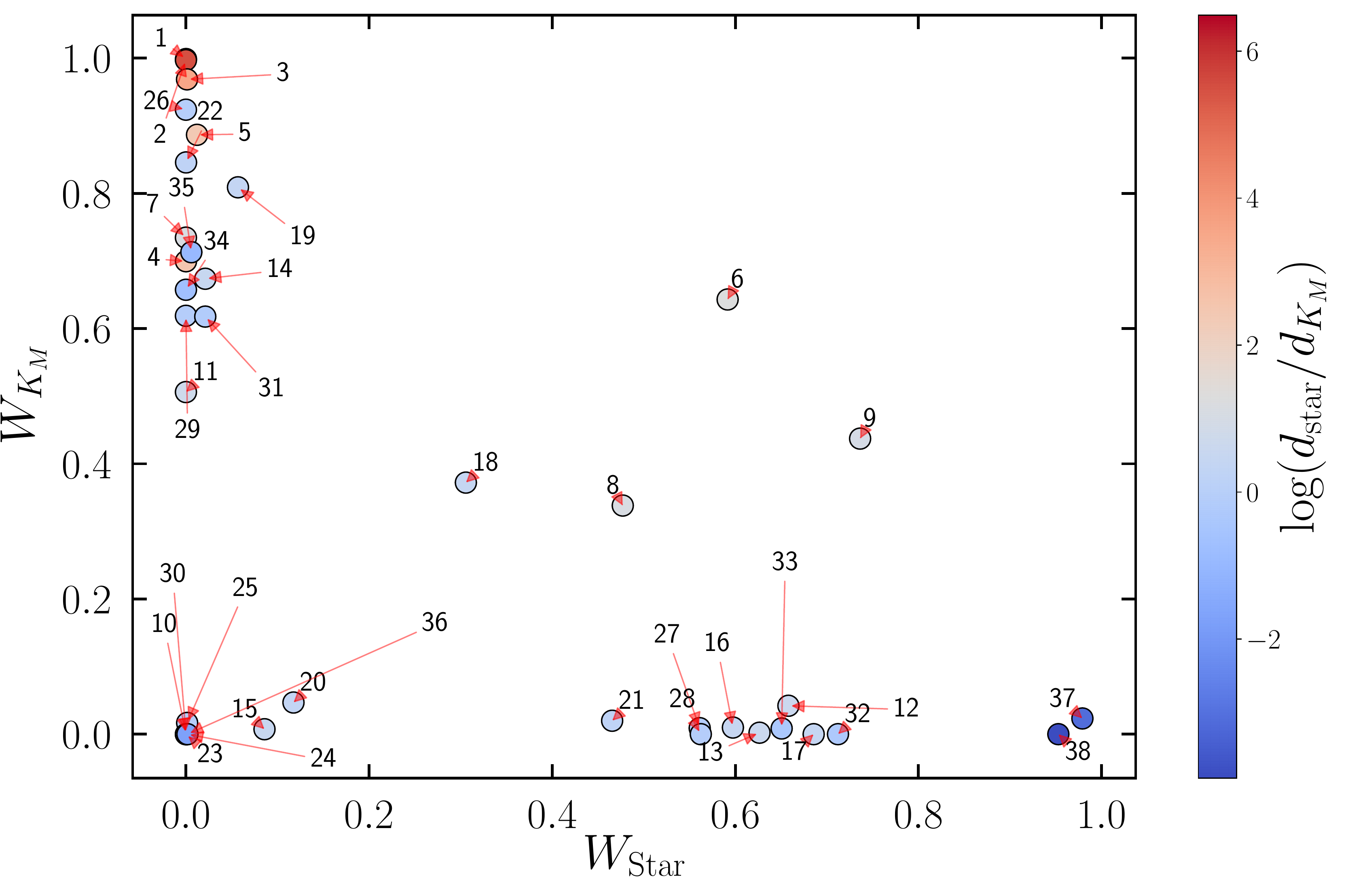}
    \caption{Scatter plot of the total PEV weight localized on the max
      $K$-core ($W_{K_M}$) versus the total PEV weight localized on the
      star graph around the hub ($W_\mathrm{Star}$) in the real networks
      considered. Points are identified by the network order in
      Table~\ref{tab:realnetworks}. The color of the symbols encodes the
      ratio $d_{star}/d_{K_M}$, see Table~\ref{tab:realnetworks}.}
      \label{WcorevsWhub}
  \end{center}
\end{figure}

In the case of real networks, for which only a single snapshot of a fixed
size is available, the direct inspection of the IPR does not provide
enough information, since a size dependence cannot be estimated as
for synthetic networks. In this case, therefore, we consider
the relative weights $W_{K_M}$ and $W_{Star}$ on the max $K$-core and on the
star formed by the hub and its leaves, respectively, which give a
measure of the total weight concentrated on each of these subgraphs.
In Fig.~\ref{WcorevsWhub} we present a scatter plot of these two
relative weights. The labels refer to the network numbers
presented in Table~\ref{tab:realnetworks}.
It turns out that most networks are actually strongly localized either
on the max $K$-core (large $W_{K_M}$, very small $W_{Star}$) or around
the hub (large $W_{Star}$, very small $W_{K_M}$).  However, the picture
is more complicated than what could be naively anticipated, and several
exceptions are evident. The color code of each point provides additional
information, representing the ratio $d_{star}/d_{K_M}$ between the relative
distances of subgraphs largest eigenvalues to the global $\Lambda_M$:
\begin{eqnarray}
  \label{eq:6}
  d_{star} &=& \frac{\Lambda_M - \Lambda_M^{(star)}}{\Lambda_M},\\
  d_{K_M} &=& \frac{\Lambda_M - \Lambda_M^{(K_M)}}{\Lambda_M}.
\end{eqnarray}
If the ratio $d_{star}/d_{K_M}$ is much larger or much smaller than $1$,
then the naive expectation is correct: the PEV is strongly localized on
the subgraph with largest eigenvalue. However, if the ratio is not very
different from 1 (i.e., in a range approximately between $0.4$ and $2$)
the naive expectation is not predictive: in some cases
$\Lambda_M^{(K_M)}> \Lambda_M^{(star)}$, yet the PEV is localized on the
star; the opposite case also occurs.  This result can be understood by
considering that the localization can occur on a densely interconnected
network subset, for which the max $K$-core is only an approximation.
Moreover, $\av{q}_{K_M}$ is in its turn only an approximation of largest
eigenvalue of the max $K$-core.  This is why Eq.~(\ref{conjecture}) is
an approximate formula for $\Lambda_M$ and, if the two quantities
appearing on the r.h.s. are not too different, one cannot strictly
deduce where the PEV is localized from which of the two quantities is
largest.

\begin{figure}[t]
  \begin{center}
    \includegraphics [width=\textwidth]{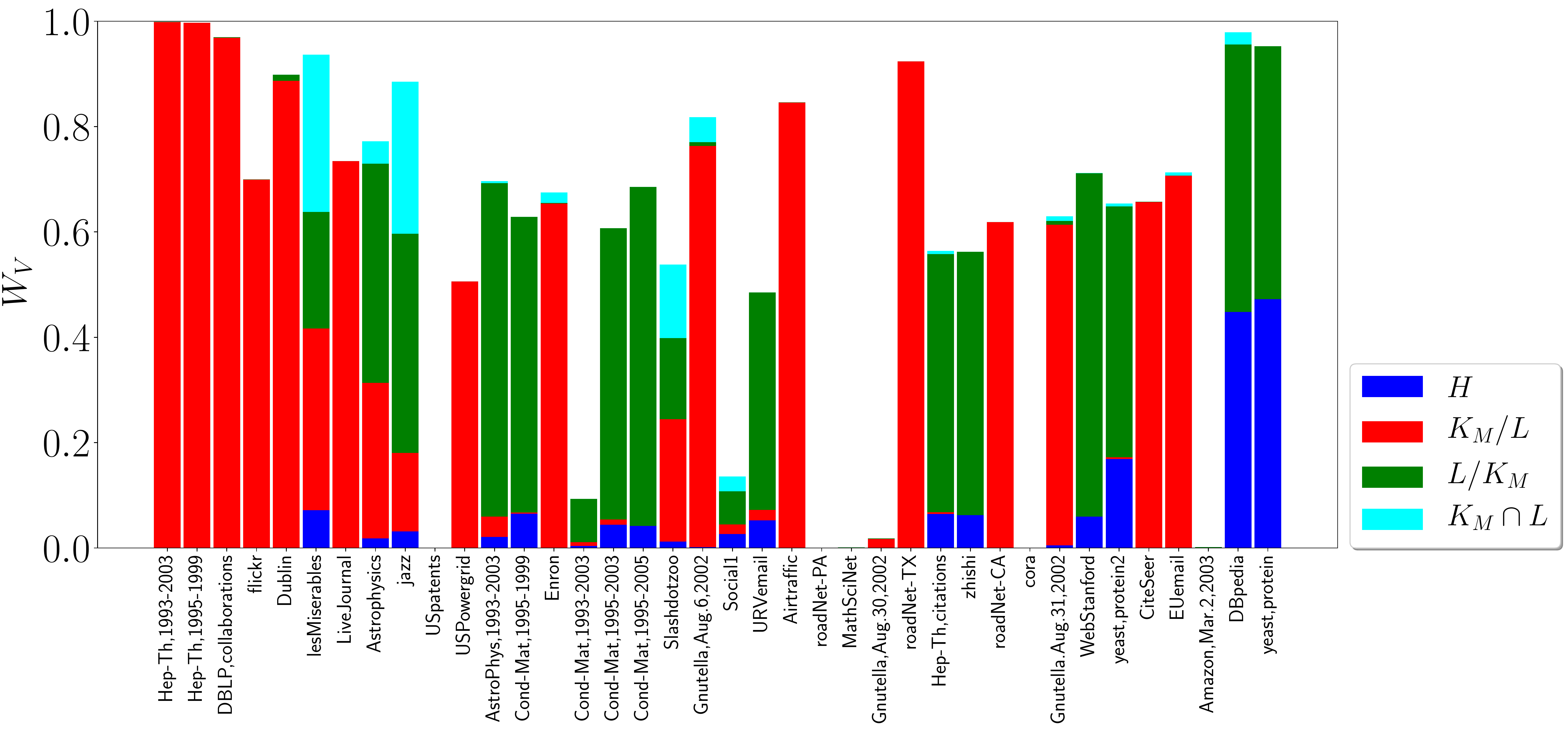}
    \caption{For each of the real networks considered, we plot the total
      PEV weight concentrated on the hub, $W_H$; on the intersection of
      the max $K$-core and of the hub leaves, $W_{K_M \cap L}$; on the
      leaves not belonging to the max $K$-core, $W_{L \setminus K_M}$; and on
      the nodes in the max $K$-core that are not leaves, $W_{K_M \setminus L}$.}
    \label{Confusion}
  \end{center}
\end{figure}

Further support to this interpretation is provided by
Fig.~\ref{Confusion}, in which we report, for each real network, the
fraction of the total weight $W_V$ localized on the hub ($H$), on the
intersection of the max $K$-core and the leaves ($K_M \cap L$), on the
leaves not belonging to the max $K$-core ($L \setminus K_M$), and on nodes
belonging in the max $K$-core that are not leaves ($K_M \setminus L$).
In the four networks having large PEV weight on both subsets, there is a
considerable weight localized on nodes shared by the two subsets; 
in other words,
some neighbors of the hub are also part of the max $K$-core: the subsets
have a strong overlap and this explains why both $W_{K_M}$ and $W_{H+L}$
are large.

Finally, a handful of networks in Fig.~\ref{WcorevsWhub} have vanishing
small weight on both subsets considered. This is a priori not
surprising: the PEV could be fully delocalized. However, it turns out
that these networks are actually localized, but on sets different from
the star or the max $K$-core; nevertheless the value of the LEV is still
approximately described by Eq.~(\ref{conjecture}).

\begin{figure}[t]
  \begin{center}
    \includegraphics [width=0.9\textwidth]{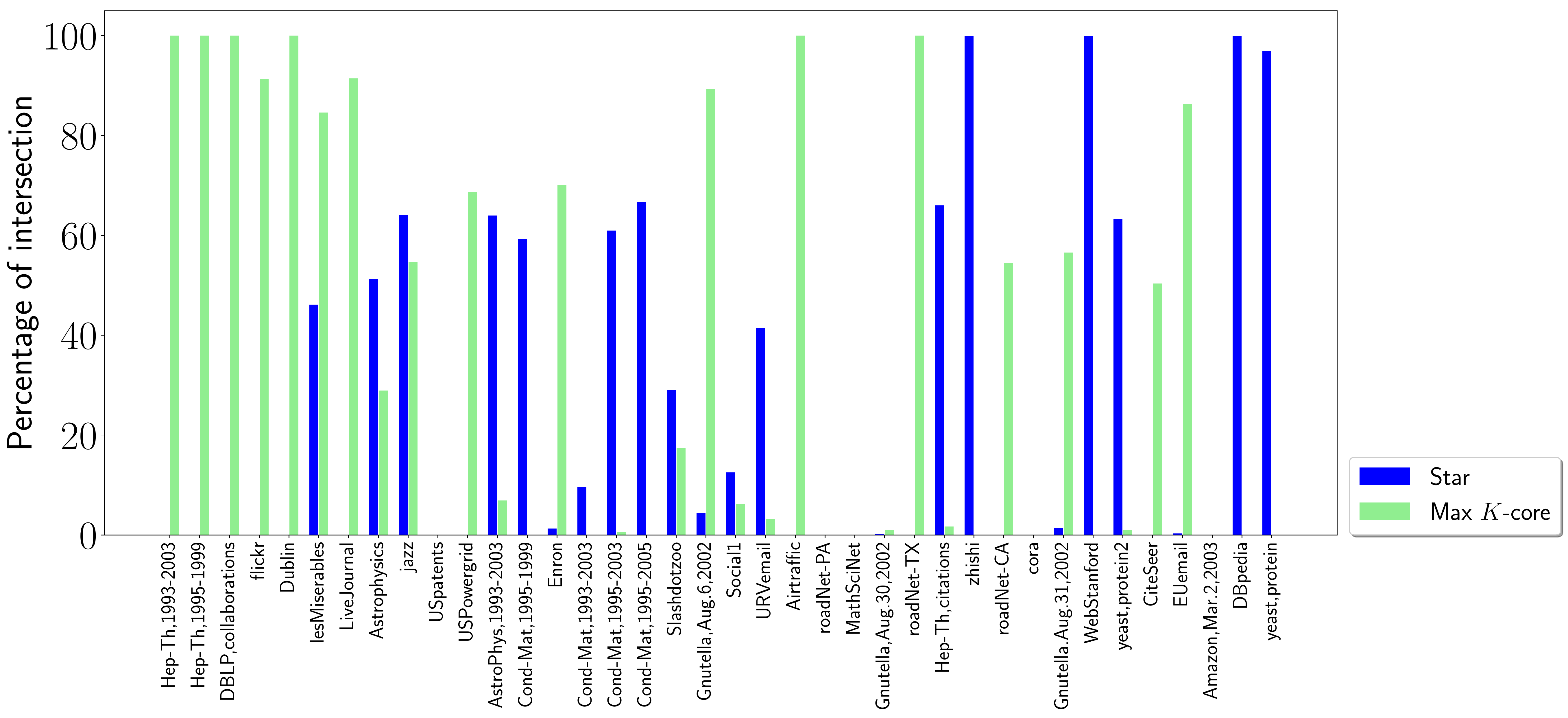}
    \caption{Intersection of the empirical localization set obtained
      from a greedy algorithm with the star and the max $K$-core
      subgraphs of the real networks considered.}
      \label{greedy}
  \end{center}
\end{figure}

This same picture can be recovered if one estimates the empirical
localization set $\mathcal{L}$ of each real network by applying a greedy
algorithm. In this approach, the PEV components $f_i^2$ are ordered in
decreasing value, and the localization set is defined by the smallest
number of nodes carrying a weight larger than or equal to a given
threshold $W_c$. In Fig.~\ref{greedy} we show the intersection of the
empirical localization set $\mathcal{L}$ obtained for a threshold
$W_c = 0.75$ with both the star and the max $K$-core subgraphs.
This figure recovers essentially the features observed in
Fig.~\ref{Confusion}: a majority of networks have an empirical
localization set that coincides essentially with the star or the max
$K$-core subgraphs, in accordance, with some exceptions with the LEV
criteria (networks in the Fig.~\ref{greedy} are ordered in
decreasing ratio $d_{star}/d_{K_M}$, see
Table~\ref{tab:realnetworks}). Some other networks, in particular
\texttt{USPatents} (10), \texttt{roadNet-PA} (23), \texttt{MathSciNet}
(24), and \texttt{cora} (30), are localized in sets with almost null
overlap with both the star and the max $K$-core. Accordingly, these
networks fall in the bottom left corner of Fig.~\ref{WcorevsWhub}.

\section{Consequences for epidemic dynamics}
\label{Consequences}

The PEV localization on different network subgraphs is an important
piece of information that can be exploited to modify and control
dynamical processes mediated by the network topology.  An immediate
example is provided by the susceptible-infected-susceptible (SIS) model
for epidemic spreading on
networks~\cite{Pastor-Satorras:2014aa}. According to this model,
individuals sit on the nodes of a network (representing the contact
pattern among them). Each individual can be either healthy but
susceptible (S) to contract the disease or infected (I) and able to
transmit the infection to his/her contacts.  For each contact between an
individual in state I and one in state S there is a rate of transmission
$\beta$, while each infected agent can spontaneously recover at rate
$\mu$. The ratio $\lambda=\beta/\mu$ is the control parameter of the
model, determining whether in the stationary state all individuals are
in state S (for $\lambda \leq \lambda_c$) or a finite fraction of them
is infected (for $\lambda > \lambda_c$).  The quenched mean-field
approach to these
dynamics~\cite{Wang03,PVM_ToN_VirusSpread,Castellano2010} predicts that
the epidemic threshold is the inverse of the adjacency matrix's LEV,
$\lambda_c = 1/\Lambda_M$. We note here that the SIS dynamics is ruled
by the spectral properties of the adjacency matrix because reinfection
events between connected nodes play a fundamental role in sustaining the
steady state characterizing the model~\cite{Castellano2012}.  In other
epidemic processes without a steady state, such as the SIR model,
reinfection events (``backtracking paths'') are not allowed, and thus
the dynamical properties in this case are described by the spectral
properties of the non-backtracking matrix
\cite{Krzakala2013,Karrer2010,Karrer2014}.

One of the crucial problems in epidemic spreading is preventive
immunization~\cite{WANG20161}: if it is possible to vaccinate only some
individuals, how should they be selected in order to minimize the
probability of an epidemic to occur and its extension?  Different strategies
have been proposed. Among them a naturally appealing one is degree-based
selection: nodes with higher degree should be vaccinated
first~\cite{PhysRevE.65.036104}.  The results presented in the previous
section suggest that the effect of immunizing high-degree individuals
can be completely different depending on where the PEV is localized.  To
show this, let us consider two networks with completely different
properties concerning PEV localization: {\tt DBpedia} (strongly
localized on the star graph around the hub) and {\tt Hep-Th,1995-1999}
(strongly localized on the max $K$-core).  In these networks, we
immunize only the node with largest degree, effectively removing it from
the network.  A first measure of the different effects of this
immunization for the two networks is the relative reduction of the LEV
$\Lambda_M$. The original networks have a LEV (see
Table~\ref{tab:realnetworks}) $\Lambda_M = 716.3$ for \texttt{DBpedia}
and $\Lambda_M = 18.04$ for \texttt{Hep-Th,1995-1999}. The resulting
networks after the removal of the hub have instead LEVs
$\Lambda'_M = 563.6$ for \texttt{DBpedia} and $\Lambda'_M = 18.04$ for
\texttt{Hep-Th,1995-1999}. Therefore, while the removal of the hub does
not affect the LEV of \texttt{Hep-Th,1995-1999}, it induces a $21\%$
decrease in \texttt{DBpedia}.

\begin{figure}[t]
  \begin{center}
    \includegraphics [width=\textwidth]{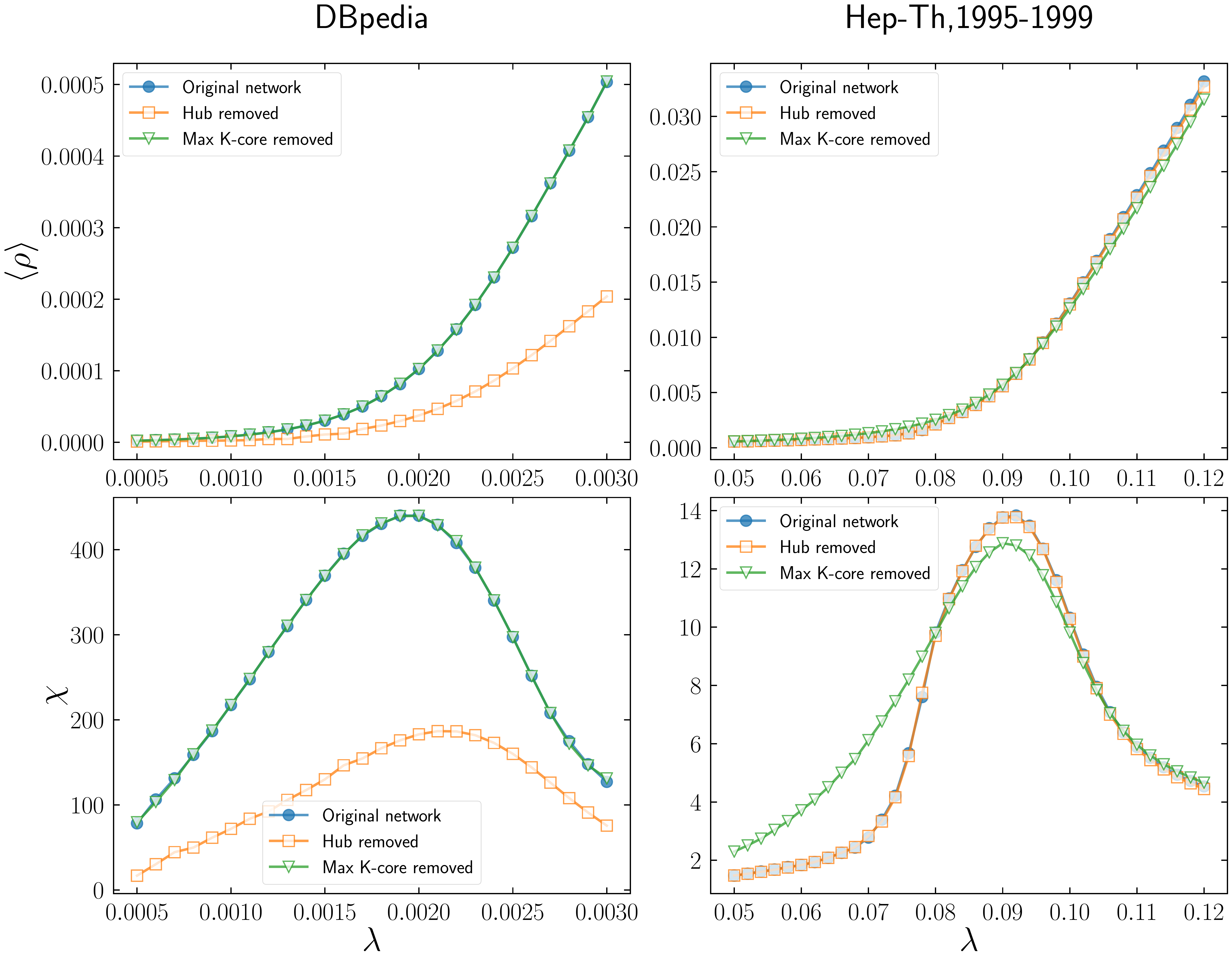}
    \caption{Results of simulations of the SIS model on the
      \texttt{DBpedia} (left column) and \texttt{Hep-Th,1995-1999}
      (right column) networks.  Top plots: density $\av{\rho}$ of nodes
      in the infected state in the quasi-stationary state.  Bottom
      plots: susceptibility
      $\chi = N(\av{\rho^2}-\av{\rho}^2)/\av{\rho}$.}
      \label{sis}
  \end{center}
\end{figure}

The different effect of the removal of a single individual on the
epidemic spreading is most clearly seen in Fig.~\ref{sis}, where we plot
the order parameter $\av{\rho}$, defined as the average number of
infected individuals in the steady state, and the susceptibility $\chi$,
defined as
\begin{equation}
  \label{eq:7}
  \chi = N \frac{\av{\rho^2} - \av{\rho}^2}{\av{\rho}},
\end{equation}
evaluated around the epidemic transition both in the presence and in the
absence of the hub. Results are obtained by applying the
quasi-stationary simulation method~\cite{Ferreira2012}.

The removal of the hub has virtually no effect in the case of
\texttt{Hep-Th,1995-1999}, where the localization occurs on the max
$K$-core.  This is expected based on the physical picture underlying
Eq.~(\ref{conjecture}): even if the hub is immunized, the dense web of
contacts among the other nodes in the max $K$-core is as effective as
before in triggering the epidemic transition.  A completely different
scenario emerges when the localization occurs around the hub, as in
\texttt{DBpedia}. The vaccination of the center of the star graph
completely destroys the local structure triggering the epidemics,
leading to a very strong reduction in the extent of the
infection. Notice that this is the effect of the removal of just one
individual out of more than 3 million. With respect to the number of
edges removed, in the case of \texttt{DBpedia} immunization of the hub
amounts to the removal of $3.7\%$ of edges, while in
\texttt{Hep-Th,1995-1999} it implies the deletion of less than $0.4\%$
of edges. Such difference in the fraction of edges removed
cannot does not account for the strong effect observed.

In the same Fig.~\ref{sis}, we present also the results obtained for SIS
dynamics on the two networks considered, when the nodes belonging to
the maximum $K$-core have been immunized. In the case of
\texttt{DBpedia}, the immunization of the max $K$-core produces
essentially no effect at all, in agreement with the picture in which the
weight of the PEV is strongly localized in the hub and its neighbors. On
the other hand, the max $K$-core immunization in
the \texttt{Hep-Th,1995-1999} network induces a noticeable variation in
the SIS dynamics, with a stronger effect at the level of the
susceptibility $\chi$. This is again in agreement with the fact the that
PEV is, in this case, localized on the max $K$-core. While the effect is
in appearance small, one must bear in mind that it is produced by the
immunization of a max $K$-core of size $19$, implying the removal of
barely $1.3\%$ of edges, which is sufficient to induce a variation of a
$20.4\%$ in the value of the LEV.

The two examples presented are the most extreme. For other networks,
with weaker localization on subgraphs, one expects less strong, yet
similar, effects.

\section{Conclusions}
\label{Conclusions}

The principal eigenvector (PEV) of the adjacency matrix
characterizing complex networks has interesting localization properties,
which are associated with the spread and containment of diseases in
networked substrates. In this work, leveraging the results of
Ref.~\cite{Castellano2017}, we have shown that the localization of the
PEV in real empirical networks is often strongly correlated with the value of
its associated largest eigenvalue (LEV), and in its turn, with the
network subgraph that dominates this LEV. In this sense, the LEV is
essentially given by the maximum between the largest eigenvalue of two
well-defined subgraphs: the star formed by node of largest degree (hub)
and its nearest neighbors, and the densely connected subgraph given by 
the maximum core in a $K$-core decomposition. When the global LEV
is very close to the LEV of one of these subgraphs, the PEV is
correspondingly strongly localized in that subgraph. This
picture applies in general, but with non negligible exceptions. 
The predictive power of our result
depends on how close the LEV is to one of the subgraph LEV's, and thus
on the distance between these subgraph LEV's. When the LEV of the whole
network is very close to the LEV of a sugbraph, strong
localization ensues. When the LEV is at an intermediate position,
different levels of localization mixing between the subgraphs are
possible. In other cases, moreover, localization can be strong, but in a
subgraph with no intersection with either the star or the max $K$-core,
depending on particular features of the network's topology.

Our results have a bearing also on the behavior of epidemic processes on
networks. Indeed, the subgraph in which the PEV is localized has been
identified with the activation set of epidemic
spreading~\cite{Castellano2012}. Here we show that the fact that the
localization subgraph can be a $K$-core, independent of the hub, renders
usual degree-based immunization strategies useless, since the
vaccination of the hub can leave almost untouched the $K$-core, where 
epidemic activation is centered. Our results open thus the path for the 
research of more global immunization strategies, taking into account the full
spectral structure of complex networks.

\section*{Appendix}
In this appendix we present data about the real-world networks considered in
the analysis.

\begin{table}[p]
  \centering
% \begin{ruledtabular}

  \begin{tabular}{ll|c|cccccccc}
  &  Network &  $d_\mathrm{star}  / d_\mathrm{K_M}$  &  $N$& $W_\mathrm{H+L}$
    &  $W_\mathrm{K_M}$  &    $\Lambda_N$  & $\sqrt{q_\mathrm{max}}$ &
    $\av{q}_\mathrm{K_M}$ & $N_\mathrm{K_M}$
                                                             \\
\hline
1 & Hep-Th,1993-2003 & 659.3 & 8638 & 3.9e-05 & 0.9987 & 31.03 & 8.062 & 31 & 32 \\
2 & Hep-Th,1995-1999 & 248.2 & 5835 & 7.129e-10 & 0.9971 & 18.04 & 7.071 & 18 & 19 \\
3 & DBLP,collaborations & 34.19 & 317080 & 0.001186 & 0.9686 & 115.8 & 18.52 & 113 & 114 \\
4 & flickr & 12.73 & 105722 & 5.479e-05 & 0.6996 & 615.6 & 73.65 & 573 & 574 \\
5 & Dublin & 11.29 & 410 & 0.01207 & 0.8866 & 23.38 & 7.071 & 21.94 & 32 \\
6 & lesMiserables & 3.591 & 77 & 0.5916 & 0.643 & 12.01 & 6 & 10.33 & 12 \\
7 & LiveJournal & 3.201 & 5189808 & 1.123e-07 & 0.7345 & 539 & 122.5 & 408.9 & 415 \\
8 & Astrophysics & 3.07 & 14845 & 0.4771 & 0.3382 & 73.89 & 18.97 & 56 & 57 \\
9 & jazz & 2.723 & 198 & 0.7363 & 0.4372 & 40.03 & 10 & 29 & 30 \\
10 & USpatents & 2.309 & 3764117 & 8.901e-06 & 2.185e-11 & 113 & 28.16 & 76.28 & 106 \\
11 & USPowergrid & 2.107 & 4941 & 1.013e-20 & 0.506 & 7.483 & 4.359 & 6 & 12 \\
12 & AstroPhys,1993-2003 & 1.873 & 17903 & 0.658 & 0.04205 & 94.43 & 22.45 & 56 & 57 \\
13 & Cond-Mat,1995-1999 & 1.834 & 13861 & 0.6264 & 0.002197 & 24.98 & 10.34 & 17 & 18 \\
14 & Enron & 1.68 & 33696 & 0.02111 & 0.6737 & 118.4 & 37.19 & 70.06 & 275 \\
15 & Cond-Mat,1993-2003 & 1.644 & 21363 & 0.08579 & 0.007321 & 37.89 & 16.7 & 25 & 26 \\
16 & Cond-Mat,1995-2003 & 1.6 & 27519 & 0.5974 & 0.009829 & 40.31 & 14.21 & 24 & 25 \\
17 & Cond-Mat,1995-2005 & 1.553 & 36458 & 0.6856 & 3.991e-10 & 51.29 & 16.67 & 29 & 30 \\
18 & Slashdotzoo & 1.522 & 79116 & 0.3057 & 0.3721 & 130.4 & 50.34 & 77.8 & 129 \\
19 & Gnutella,Aug.6,2002 & 1.499 & 8717 & 0.05683 & 0.8089 & 22.38 & 10.72 & 14.61 & 175 \\
20 & Social1 & 1.43 & 67 & 0.1174 & 0.04684 & 5.591 & 3.317 & 4 & 5 \\
21 & URVemail & 1.264 & 1133 & 0.4655 & 0.01989 & 20.75 & 8.426 & 11 & 12 \\
22 & Airtraffic & 1.258 & 1226 & 0.0001306 & 0.8458 & 9.209 & 5.831 & 6.523 & 107 \\
23 & roadNet-PA & 1.219 & 1087562 & 2.651e-38 & 1.89e-14 & 4.42 & 3 & 3.255 & 916 \\
24 & MathSciNet & 1.143 & 332689 & 0.001357 & 2.804e-07 & 36.09 & 22.27 & 24 & 25 \\
25 & Gnutella,Aug.30,2002 & 0.9298 & 36646 & 0.001329 & 0.01666 & 12.93 & 7.416 & 7 & 14 \\
26 & roadNet-TX & 0.9287 & 1351137 & 9.823e-48 & 0.9238 & 4.906 & 3.464 & 3.353 & 1491 \\
27 & Hep-Th,citations & 0.9166 & 27400 & 0.5611 & 0.008903 & 111.3 & 49.68 & 44.08 & 52 \\
28 & zhishi & 0.8983 & 372840 & 0.5623 & 1.182e-11 & 1008 & 356.5 & 282.7 & 449 \\
29 & roadNet-CA & 0.8905 & 1957027 & 1.499e-09 & 0.6189 & 4.638 & 3.464 & 3.32 & 4454 \\
30 & cora & 0.8594 & 23166 & 4.556e-05 & 4.623e-06 & 31.5 & 19.42 & 17.44 & 25 \\
31 & Gnutella.Aug.31,2002 & 0.8357 & 62561 & 0.0211 & 0.6177 & 13.18 & 9.747 & 9.072 & 1004 \\
32 & WebStanford & 0.749 & 255265 & 0.7121 & 2.445e-05 & 448.1 & 196.5 & 112.2 & 387 \\
33 & yeast,protein2 & 0.6756 & 2172 & 0.6506 & 0.008435 & 21.1 & 14.66 & 11.57 & 14 \\
34 & CiteSeer & 0.5007 & 365154 & 2.891e-05 & 0.6574 & 58.08 & 41.7 & 25.37 & 1850 \\
35 & EUemail & 0.3844 & 224832 & 0.005901 & 0.7129 & 102.5 & 87.38 & 63.12 & 292 \\
36 & Amazon,Mar.2,2003 & 0.1807 & 262111 & 0.001668 & 3.41e-05 & 23.55 & 20.49 & 6.643 & 286 \\
37 & DBpedia & 0.04497 & 3915921 & 0.9791 & 0.02326 & 716.3 & 685.3 & 27.89 & 70 \\
38 & yeast,protein & 0.02039 & 1458 & 0.9528 & 4.17e-06 & 7.535 & 7.483 & 5 & 6 \\

  \end{tabular}
%\end{ruledtabular}
  \caption{Topological and spectral properties of the real networks
    considered: Ratio of the distance between the actual largest
    eigenvalue $\Lambda_M$ and the LEV's of the relevant subgraphs (star
    centered at the hub and max $K$-core),
    $d_\mathrm{star} / d_\mathrm{K_M}$; network size $N$, relative
    weight of the PEV on the star centered in the hub, $W_\mathrm{H+L}$;
    relative weight of the PEV on the max $K$-core, $W_\mathrm{K_M}$;
    largest eigenvalue $\Lambda_M$; square root of the maximum degree,
    $\sqrt{q_\mathrm{max}}$; average degree of the max $K$-core,
    $\av{q}_\mathrm{K_M}$; size of the max $K$-core, $N_\mathrm{K_M}$.}
  \label{tab:realnetworks}
\end{table}

\end{document}